\documentclass[aps,10pt,prl,twocolumn,groupedaddress]{revtex4}

\usepackage{graphicx}
\usepackage[caption=false]{subfig}
\usepackage{amsmath}
\usepackage{bbold}
\usepackage{latexsym}

\begin{document}

%\preprint{}

\title{Experimental constraints and a possible quantum Hall state at $\nu =5/2$}


\author{Guang Yang and D. E. Feldman}
%\email[]{}
%\thanks{}
\affiliation{Department of Physics, Brown University, Providence, Rhode Island 02912, USA}


\date{\today}

\begin{abstract}

Several topological orders have been proposed to explain the quantum Hall plateau at $\nu=5/2$. The observation of an upstream neutral mode on the sample edge supports the non-Abelian anti-Pfaffian state. On the other hand, tunneling experiments favor the Halperin 331 state which exhibits no upstream modes. No proposed ground states agree with both types of experiments. We find a topological order, compatible with the results of both experiments. That order allows both finite and zero spin polarizations. It is Abelian but its signatures in Aharonov-Bohm interferometry can be similar to those of the Pfaffian and anti-Pfaffian states. 

\end{abstract}


% insert suggested PACS numbers in braces on next line \pacs{}
%\keywords{}

\maketitle

% References should be done using the \cite, \ref, and \label command

Fractional quantum Hall effect (QHE) exhibits remarkably rich phenomenology. More than 70 filling factors have been discovered. Some of them are well understood but many are not. 
In particular, the nature of the fragile states in the second Landau level remains a puzzle.

The quantum Hall plateaus at the filling factors \cite{7} $\nu=5/2$ and $\nu=7/2$ are particularly interesting. Almost all known filling factors have odd denominators. Such quantum Hall states can be explained in a natural way within the Haldane-Halperin hierarchy \cite{wen} and the composite fermion picture \cite{qhe2}. Even-denominator filling factors require additional ideas. It was argued that electrons form pairs \cite{pair} at $\nu=5/2$, {\it i.e.}, the $5/2$ state is a topological superconductor. Paring implies that the lowest-charge quasiparticles carry one quarter of an electron charge \cite{2,7}. This was indeed observed in several 
experiments \cite{9,willett09,21}. At the same time, the nature of pairing remains an open problem.

The investigations of the $\nu=5/2$ QHE liquid have focused on its topological order which is robust to small variations of sample parameters \cite{wen}. 
This led to a striking proposal of non-Abelian statistics \cite{2,7}. In contrast to ordinary fermions, bosons and Abelian anyons, systems of non-Abelian quasiparticles possess numerous degenerate ground states at fixed quasiparticle positions. This may be useful for quantum computing \cite{7}. Theoretically proposed non-Abelian Pfaffian, anti-Pfaffian, $SU(2)_2$ and anti-$SU(2)_2$ states have attracted much interest as possible candidates to explain the QHE plateau at $\nu=5/2$ (for a review of the proposed states see Refs. \onlinecite{Overbosch} and \onlinecite{yf13}). At the same time, one can also construct Abelian states with the same filling factor, such as the Halperin 331, $K=8$ and anti-331 states \cite{Overbosch,yf13}.

Most above-mentioned states were invented before experimental information beyond the existence of the $5/2$ QHE plateau and the value of its energy gap became available. 
This made it impossible to select the correct theory of the $5/2$-liquid.
The last few years have seen considerable accumulation of the new experimental results \cite{9,willett09,21,willett10,36,mstern10,rhone11,tiemann12,mstern12,chickering13,20,baer}. They provide tight constraints on the topological order at $\nu=5/2$. We argue that all previously proposed ground state wave functions are excluded by those constraints. To explain the $5/2$ plateau we
propose a different topological order that satisfies the experimental constraints and thus is a serious candidate to solve the $5/2$ puzzle.

Since quasiparticle statistics is defined in terms of particles moving around each other, the smoking-gun probe of topological order is interferometry \cite{7}. 
It was argued that some of the Aharonov-Bohm interferometry results are compatible with the non-Abelian Pfaffian state \cite{bishara09,willett10}. However, the 331 state may show similar interferometric signatures \cite{fp331}
 and a more sophisticated Mach-Zhender interferometer may be necessary to distinguish it from the Pfaffian state \cite{16}. 
Spin polarization data are controversial: optical experiments \cite{mstern10,rhone11} were interpreted as a sign of zero polarization while resistively-detected NMR points
\cite{tiemann12,mstern12} at $100\%$ polarization. 
Thermoelectric response \cite{chickering13} shows qualitative agreement with a non-Abelian state but an Abelian state may exhibit similar behavior, if it has different types of quasiparticles of close or equal energy.

Several groups \cite{21,20,baer} performed tunneling experiments which measured the quasiparticle current through a narrow constriction. At low temperatures the zero-bias conductance scales as $T^{2g-2}$, where the exponent $g$ depends on the topological order and is universal in the absence of long-range interactions \cite{wen}. Theory predicts $g=1/2$ in the anti-Pfaffian and $SU(2)_2$ states and $g=3/8$ in the 331 state \cite{Overbosch,yf13}. All existing predictions \cite{yf13} for other states are either above $1/2$ or below $3/8$. In the earliest experiment \cite{20}, the best fit for $g$ at the fixed charge $e^*=e/4$ of the tunneling quasiparticle was $g=0.45$. This was  interpreted initially as a signature of the anti-Pfaffian or $SU(2)_2$ state. Subsequent experiments \cite{21,baer} in other sample geometries produced $g$ between $0.37$ and $0.42$ as the best fits at fixed $e^*=e/4$. This supports the case for the 331 state. It was argued that the measured exponents are affected by long-range electrostatic forces \cite{yf13}. Their effect depends on the sample geometry and in all cases increases the observed $g$. Thus, all tunneling data are compatible with the 331 state \cite{yf13}.

At the same time, the 331 state is incompatible with the observation \cite{36} of an upstream neutral mode.
This means that no proposed ground state wave function fits all existing data. Below we identify a different ground state that agrees with the existing experiments.

We propose that the $5/2$ liquid exhibits Halperin's 113 topological order. This possibility was addressed in the numerical study \cite{x} but did not receive much further attention. Moreover, it was pointed out \cite{gail} that the wave function from Ref. \onlinecite{x} is unsuitable to explain the $5/2$ plateau. The usual Halperin $nnm$ wave function is
\begin{align}
\label{D1}
\Psi_{\rm usual}=\Pi_{k<l}(z_k-z_l)^n\Pi_{\alpha<\beta}(w_\alpha-w_\beta)^n\times \nonumber\\
\Pi_{k,\alpha}(z_k-w_\alpha)^m \exp(-\frac{1}{4l_B^2}\sum [|z_l|^2+|w_\alpha|^2]),
\end{align}
where $l_B$ is the magnetic length, $z_k=x_k+iy_k$ and $w_\alpha=x_\alpha+i y_\alpha$ are the positions of the two flavors of electrons, the simplest possibility for the flavor degree of freedom being spin. The plasma analogy shows that such wave function exhibits phase separation \cite{gail} into single-flavor regions at $n<m$. This seems to invalidate the possibility of the 113 topological order. On the other hand, the above phase-separation argument also applies to the 112 order, believed to describe the spin-singlet state \cite{112}
 at low magnetic fields at the filling factor $2/3$. The apparent inconsistency is resolved by noting that the same topological order can be encoded in many wave functions and one can find an $nnm$ wave function, free of pathology at $n<m$.  In particular, the following $nnm$ wave function with $n<m$ can be built by analogy with Laughlin's quasiparticle construction and was argued \cite{112} to describe the spin-singlet $2/3$ state at $n=1$ and $m=2$:

\begin{align}
\label{D2}
\Psi= \hat P\exp(-\frac{1}{4l_B^2}\sum [|z_l|^2+|w_\alpha|^2]) \times\nonumber\\
\Pi_{k<l}(\partial_{z_k}-\partial_{z_l})^{m-n}\Pi_{\alpha<\beta}(\partial_{w_\alpha}-\partial_{w_\beta})^{m-n}\times\nonumber\\ \Pi_{k<l}(z_k-z_l)^m
\Pi_{\alpha<\beta}(w_\alpha-w_\beta)^m
\Pi_{k,\alpha}(z_k-w_\alpha)^m,
\end{align}
where the operator $\hat P$ takes care about appropriate antisymmetrization with respect to the flavor degree of freedom. The same expression (\ref{D2}) with $n=1$ and $m=3$ describes the 113 order. At the same time, we would like to emphasize that below we focus on the observable consequences of the 113 topological order and not a particular wave function choice. Indeed, a highly symmetric wave function (\ref{D2}) might not be a good description for a realistic disordered system.

The wave function (\ref{D2}) describes electrons at the filling factor $1/2$. In addition to the second Landau level at $\nu=1/2$, the first Landau level is filled in the $5/2$ state. 

The topological properties of the 113 order are encoded in the $K$-matrix \cite{wen}

\begin{equation}
\label{D3}
K =
\left( \begin{array}{cc}
1 & 3 \\
3 & 1 \end{array} \right)
\end{equation}
and the charge vector ${\bf q}=(1,1)$. The standard formalism \cite{wen} shows that all excitations are built from two flavors of quasiparticles, represented by the vectors ${\bf l}_1=(1,0)$ and ${\bf l}_2=(0,1)$, and carrying the charge 

\begin{equation}
\label{D4}
e^*=e{\bf q}K^{-1}{\bf l}_{1,2}^T=e/4
\end{equation}
in agreement with the experiment. The mutual statistics of the two particle flavors is described by the phase, accumulated by a quasiparticle of one flavor  after it makes a full circle around a quasiparticle of the other flavor:

\begin{equation}
\label{D5}
\theta_{12}=2\pi {\bf l}_1 K^{-1}{\bf l}_2^T=3\pi/4.
\end{equation}
The statistics of two identical particles is given by the phase, accumulated when they exchange their positions:

\begin{equation}
\label{D6}
\theta_{11}=\theta_{22}=\pi {\bf l}_1 K^{-1} {\bf l}_1^T=-\pi/8.
\end{equation}

The simplest interpretation of the two quasiparticle flavors implies a spin-unpolarized state, where excitations with two different spin projections are allowed. The 113 order is also possible in a spin-polarized system. 
For example, one can rewrite Eq. (\ref{D3}) as

\begin{equation}
\label{DA1}
K =W^T K' W; K'=\left( \begin{array}{cc}
1 & 2 \\
2 & -4 \end{array} \right);
W=\left( \begin{array}{cc}
1 & 1 \\
0 & 1 \end{array} \right).
\end{equation}
The matrix $K'$ describes the same 113 topological order and can be interpreted within the hierarchical construction for spin-polarized electrons: a condensate of charge-$2e$ quasiholes forms on top of the integer QHE state.

We now turn to the edge physics \cite{wen}. The edge action is

\begin{align}
\label{D7}
L = -\frac{\hbar}{4\pi}\int dx dt{\large\{}  \sum_{i,j=1,2} K_{ij}\partial_{t}\phi_{i}\partial_{x}\phi_{j}+
\nonumber\\
 \sum_{j=I1,I2}\partial_t\phi_{j}\partial_x\phi_j +\sum_{i,j=1,2,I1,I2} V_{ij}\partial_{x}\phi_{i}\partial_{x}\phi_{j} {\large\}},
\end{align}
where the fields $\phi_{1},\phi_2$ describe the fractional QHE edge modes and $\phi_{I1},\phi_{I2}$ describe the two integer edge channels. 
By introducing the charged mode $\phi_\rho=\phi_1+\phi_2$ and the neutral mode $\phi_n=\phi_1-\phi_2$
we can rewrite the action in the form

\begin{align}
\label{D8}
L =  -\frac{\hbar}{4\pi}\int dtdx  [ 2\partial_{t}\phi_{\rho}\partial_{x}\phi_{\rho}-\partial_{t}\phi_{n}\partial_{x}\phi_{n} +\nonumber\\ 
2v_{\rho} (\partial_{x}\phi_{\rho})^{2} 
 +v_{n} (\partial_{x}\phi_{n})^{2}+2v_{\rho n}\partial_{x}\phi_{\rho}\partial_{x}\phi_{n}] \nonumber\\
-\frac{\hbar}{4\pi}\int dt dx [\sum_{i=I1,I2}(\partial_t\phi_i\partial_x\phi_i
+v_i(\partial_x \phi_i)^2)+\nonumber\\
2u_{12}\partial_x\phi_{I1}\partial_x\phi_{I2}+2\sum_{i=\rho,n;j=I1,I2}w_{ij}\partial_x\phi_i\partial_x\phi_j] ,
\end{align}
where the charge density in the fractional edge channels $\rho_F=e\partial_x\phi_\rho/(2\pi)$ and the charge density in the integer channels is $\rho_I=e(\partial_x\phi_{I1}+\partial_x\phi_{I2})/(2\pi)$. In what follows we will ignore the integer edge channels and concentrate on the first two lines in the action (\ref{D8}). The integer channels have little effect on our results as discussed in Supplemental material \cite{S}.

The minus sign in front of $\partial_{t}\phi_{n}\partial_{x}\phi_{n}$ signifies the existence of an upstream neutral mode in agreement with the experiment. We now turn to the quasiparticle tunneling. The most relevant quasiparticle operators create excitations of charge $e/4$ and have the form $O_{1,2}=\exp(i\phi_{1,2})=\exp (i[\phi_\rho\pm\phi_n]/2)$. As a starting point, we consider the flavor-symmetric situation with the symmetry $\phi_1\rightarrow\phi_2$,
$\phi_2\rightarrow\phi_1$. In such case $v_{\rho n}=0$ in Eq. (\ref{D8}) and the scaling dimensions of the operators $O_{1,2}$ are identical and equal $\Delta=3/16$. Weak quasiparticle tunneling between two edges at the point $x=0$ is described by the contribution to the action \cite{wen}

\begin{equation}
\label{D9}
L_T=\int dt \sum_{i=1,2}\Gamma_i O^{(t)}_i(x=0)O^{(b)\dagger}_i(x=0)+{\rm h.c.},
\end{equation}
where the indices $t$ and $b$ refer to the top and bottom edges. The low-temperature tunneling conductance $G(T)$ can be estimated \cite{wen} by performing the renormalization group procedure up to the energy scale $E\sim T$ and setting 
$G(T) \sim \Gamma^2(T)$. Thus, $G(T)\sim T^{2g-2}$, where $g=3/8$ in good agreement with the data.

At the same time, there is no reason for precise flavor symmetry. A nonzero $v_{\rho n}$ changes the above result. The scaling dimensions of the operators $O_{1,2}$ become different and correspond to two contributions to the conductance
$G_{2,1}(T)\sim T^{2g_{\pm}-2}$ with

\begin{equation}
\label{D10}
g_{\pm}=\frac{1}{\sqrt{1-c^2}}\left(\frac{3}{8}\pm\frac{c}{2\sqrt{2}}\right),
\end{equation}
where $c=\sqrt{2}v_{\rho n}/(v_\rho+v_n)$. This might suggest that the tunneling conductance is nonuniversal and no meaningful comparison with the experiment is possible. 
However, we show below that a theory, based on the 113 order, does predict the scaling of the conductance with $g$, close to $3/8$, even without the flavor symmetry. In contrast to the  famous spin-polarized $2/3$ state \cite{38} our explanation of universality does not involve disorder on the edge.

Even arbitrarily weak disorder guarantees proper quantization of the quantum Hall conductance $G=\nu e^2/h$ in a bar geometry in a system with upstream modes, if the edges are long enough. At the same time, the tunneling conductance only depends on what happens 
within a thermal length from the tunneling contact. Hence, the low-temperature tunneling
conductance is affected by disorder only if disorder is relevant in the renormalization group sense \cite{38}.
 In our problem, disorder is responsible for electron tunneling between the two fractional modes $\phi_1$ and $\phi_2$. The corresponding contribution to the action $L_D=\int dt dx \left\{\zeta(x)\exp[2i\phi_n(x)]+{\rm h.c.}\right\}$, where $\zeta(x)$ is a random complex number. One can check that $L_D$ is always irrelevant. Thus, disorder does not lead to a universal conductance scaling and Eq. (\ref{D10}) applies. The explanation for the observed $g\approx 3/8$ is different: we argue that $c\ll 1$.

\begin{figure}[b]
\centering
\includegraphics[width=1.5in]{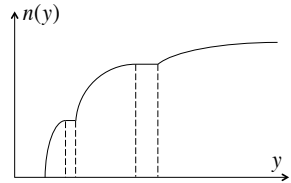}
\caption{Density profile of a 2D electron gas in a magnetic field. The density is constant in narrow incompressible strips.}
\label{BBB}
\end{figure}

We wish to estimate $v_n$, $v_\rho$ and $v_{\rho n}$ (\ref{D8}) in the tunneling experiments \cite{21,20,baer}. In all those experiments, the edges are defined by top gates. As observed in Ref. \onlinecite{csg}, the charge density profile in such situation is almost the same as in the absence of the magnetic field. The edge consists of several compressible strips, separated by narrow incompressible strips of fixed charge density (Fig. 1). The widths of the compressible strips and their distances from the gates depend on the filling factor and other details and are estimated to be between hundreds nm and a few $\mu$m \cite{csg}. The widths of the gates and their distances from the electron gas are within the same range. This gives us an estimate of the distance between the edge states and the gates. In the simplest picture, the widths of various edge channels can be estimated from the widths of the compressible strips. Quantum localization modifies such picture \cite{csg}. We expect that an edge channel is located within a compressible region between two incompressible strips with filling factors $\nu_1<\nu_2$. The part of the compressible strip on one side of the edge channel should be understood as an incompressible QHE liquid with the filling factor $\nu_2$ and localized quasiholes. The part 
of the compressible strip on the other side of the channel should be understood as an incompressible QHE liquid of the filling factor $\nu_1$ with localized quasiparticles. The width $a$ of the edge channel depends on the localization length and is less than the total width of the compressible strip. We expect $a>l_B$. A localization length $<l_B$
would mean that disorder is too strong for QHE correlations to exist.

Let us now estimate $v_\rho$. If the distance from the edge to the gate is comparable to $a$ then the energy cost of the average linear charge density $\rho=e\partial_x\phi_\rho/(2\pi)$ in a region of size $a\times a$ is

\begin{equation}
\label{D11}
\delta E\sim (a\rho)^2/(\epsilon a),
\end{equation}
 where $\epsilon$ is the dielectric constant. This energy cost enters the action (\ref{D8}) as $a\hbar v_\rho(\partial_x\phi_\rho)^2/(2\pi)$. Hence, $v_\rho\sim e^2/(\hbar\epsilon)$. The velocity $v_\rho$ increases by a factor of $\ln(d/a)$, if the edge width $a$ is much smaller than the distance $d\gg l_B$ from the gate \cite{ag}. This comes from the energy cost of the interaction between the sections of the edge at the distances $l$, $d>l>a$.

We expect that the neutral mode runs at the same place as the charged mode and the excitations of the neutral mode redistribute the two electron flavors without changing the overall charge density beyond the magnetic length $l_B$. Thus, the neutral mode only participates in short-range interaction of radius $l_B$.
We find $v_n$ by estimating the energy cost of the disbalance $\rho_n=\rho_1-\rho_2$ between the charge densities of the two flavors. We get an estimate, similar to Eq. (\ref{D11}), but with an additional factor $l_B/a$ to account for the short-range character of the interaction \cite{footnote}.
This is similar to the calculation of the charge velocity in the presence of the top gate \cite{topgate}, where the interaction radius is set by the distance to the gate. Thus,
$v_n\sim v_{\rho n}\sim \frac{l_B}{a[1+\ln(d/a)]}v_\rho\ll v_\rho$. Hence, $c\ll 1$ and $g\approx 3/8$. The latter conclusion is not affected by the integer edge channels \cite{S}. Indeed, due to spacial separation, the interaction of the neutral mode with the integer channels is weaker than its interaction with the fractional charged mode.

Our physical picture differs from the simplest picture of the charged and neutral channels in a very clean $\nu=2$ system. There, two spin channels correspond to two wide compressible strips, separated by an incompressible region.
Nevertheless, even at $\nu=2$ one expects the charged mode to be much faster than the neutral mode \cite{G13}. This is the only thing that matters for our estimate of $c$. Besides, with two contra-propagating channels, the generalization of the $\nu=2$ picture for $\nu=5/2$ would imply two wide compressible regions, one with $2<\nu<\nu_{\rm max}$  and the other with $5/2<\nu<\nu_{\rm max}$, and an incompressible strip with $\nu=\nu_{\rm max}>5/2$ in the middle. Such non-monotonous charge distribution with $\nu>5/2$
in an area of width $s\gg l_B$ differs significantly from the charge distribution in the absence of the magnetic field and is unlikely.

\begin{figure}[b]
\centering
\includegraphics[width=1.5in]{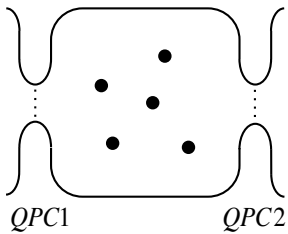}
\caption{Aharonov-Bohm interferometer. Quasiparticles move along the edges and tunnel between the edges at the quantum point contacts QPC1 and QPC2. Several quasiparticles are localized between the edges.}
\label{BBB}
\end{figure}

What are the signatures of the 113 state in an Aharonov-Bohm interferometer \cite{7}, Fig. 2? We consider two possibilities: 1) only one quasiparticle flavor can tunnel through the tunneling contacts in Fig. 2; 2)
the flavor-symmetric situation: tunneling amplitudes are identical for both flavors. The realistic situation is likely in between. In the first case the current through the interferometer changes periodically as a function of its area with the period, corresponding to the additional magnetic flux $\Phi_{1/4}=hc/e^*=4hc/e$ through the device. In the second case, we need to add two periodic patterns of period $\Phi_{1/4}$ due to the two flavors. Their phase difference $\Delta \theta$ depends on the numbers $n_{1,2}$ of the localized quasiparticles of the two flavors inside the interferometer. One finds $\Delta\theta=(2\theta_{11}-\theta_{12})(n_1-n_2)=\pi (n_1+n_2){\rm~mod~2\pi}$. Hence, the two interference patterns cancel, if an odd number of quasiparticles are localized in the device. The same behavior is expected for the Pfaffian state \cite{7} and the flavor-symmetric 331 state \cite{fp331}. 
At the same time, a more complex Mach-Zehnder setup is known to unambiguously distinguish the 331 and Pfaffian states \cite{16} and may help probe the 113 topological order.

All states, reviewed in Ref. \onlinecite{yf13}, exhibit universal tunneling transport with $g={\rm integer}/8$. According to Eq. (\ref{D10}), the 113 state is different and a very precise tunneling experiment will show $g=g_-$, where
$2/8<g_-<3/8$. Another unique signature of the 113 state is the existence of two contributions to the tunneling current with the exponents $g_-\approx g_+$ (\ref{D10}) such that $g_- + g_+=3/4$ to the second order in the small parameter $c$.

In contrast to our findings, numerical investigations of small model systems favor the Pfaffian and anti-Pfaffian states (see, e.g., Refs. \onlinecite{morf98,dimov08,feiguin2009,biddle}). At the same time, existing numerical results have a number of limitations. For example, Landau-level mixing \cite{wojs2010,peterson2013} was ignored in many studies. Numerical predictions for the energy gap are many times higher than the experimental 
findings \cite{morf2002,morf2003}. 
This may be due to disorder \cite{morf2003,dAmbrumenil2011} but no attempts have been made to include disorder in numerical simulations. Taking into account small energy differences \cite{biddle} between trial wave-functions, corresponding to different topological orders at $\nu=5/2$, one cannot make definite conclusions from numerics alone about the nature of the $5/2$ state in a realistic disordered system. Only experiment can solve the $5/2$ puzzle.

In conclusion, we propose a topological order whose properties agree with the existence of an upstream neutral mode and the observed behavior in tunneling experiments at $\nu=5/2$. No other proposed state fits with the existing body of the experimental facts. Certainly, more experiments are needed before one can   conclusively establish the nature of the $5/2$ QHE liquid.  
In particular, the confirmation of the upstream neutral mode \cite{36} with a different method is desirable. Further experiments would strengthen the conclusion \cite{21,20,baer} that the measured tunneling exponents are determined by the physics at the low-energy fixed point.
Meanwhile the 113 state should be taken as a serious candidate to explain the even-denominator QHE at $\nu=5/2$.

We thank A. Kitaev and A. M. M. Pruisken for useful discussions. This work was supported by the NSF under Grant No. DMR-1205715.

\appendix
\section{Supplemental material for ``Experimental constraints and a possible quantum Hall state at $\nu =5/2$'':
The effect of the integer modes.}

In this Supplemental material we clarify the effect of the integer QHE edge channels on the tunneling exponent $g$. There are two integer channels on the edge: the charged mode $\phi_c$ and the spin mode. We will consider a simple model in which the spin mode does not interact with the other modes and hence can be ignored. We will also neglect the interaction between the fractional QHE neutral mode $\phi_n$ and the integer charged mode $\phi_c$ and will only include the interaction of the neutral mode with the fractional QHE charged mode $\phi_\rho$ and the interaction between the integer and fractional charged modes. Thus, we consider the following edge action:

\begin{align}
\label{DS1}
L=-\frac{\hbar}{4\pi}\int dx dt
{\big [}
2\partial_t\phi_\rho\partial_x\phi_\rho+2v_\rho(\partial_x\phi_{\rho})^2\nonumber\\
-\partial_t\phi_n\partial_x\phi_n+v_n(\partial_x\phi_n)^2+\frac{1}{2}\partial_t\phi_c\partial_x\phi_c+\frac{v_c}{2}(\partial_x\phi_c)^2\nonumber\\
+2v_{\rho n}\partial_x\phi_n\partial_x\phi_\rho+2v_{\rho c}\partial_x\phi_\rho\partial_x\phi_c
{\big ]},
\end{align}
where we estimate the velocities of the charged modes as $v_\rho\sim\frac{e^2}{\epsilon\hbar}(1+\ln\frac{d}{a})$ and $v_c\sim\frac{e^2}{\epsilon\hbar}(1+\ln\frac{d'}{a'})$ with $a'$ and $d'$ being the  width of the integer edge and its distance from the gate. We expect $v_{\rho n}\sim v_n\ll v_\rho$, $v_n\le v_c$  and $v_{\rho c}\le v_\rho, v_c$.

We first diagonalize the action part that does not depend on $\phi_n$. This is achieved by introducing new variables $\theta_{1,2}$ such that 

\begin{align}
\label{DS2}
\phi_\rho=\frac{1}{\sqrt{2}}(\theta_1 \cos\alpha+\theta_2\sin\alpha);\nonumber\\
\phi_c=\sqrt{2}(-\theta_1\sin\alpha+\theta_2\cos\alpha),
\end{align}
where

\begin{align}
\label{DS3}
\cos\alpha=\sqrt{\frac{1}{2}+\frac{1}{2\sqrt{1+\frac{4v_{\rho c}^2}{(v_\rho-v_c)^2}}}};\nonumber\\
\sin\alpha=\sqrt{\frac{1}{2}-\frac{1}{2\sqrt{1+\frac{4v_{\rho c}^2}{(v_\rho-v_c)^2}}}}.
\end{align}
The action assumes the form

\begin{align}
\label{DS4}
L=-\frac{\hbar}{4\pi}\int dx dt{\big [}
-\partial_t\phi_n\partial_x\phi_n+v_n(\partial_x\phi_n)^2+\nonumber\\
\sum_{k=1}^2\partial_x\theta_k(\partial_t\theta_k+v_k\partial_x\theta_k)+\nonumber\\
\sqrt{2}v_{\rho n}\partial_x\phi_n(\cos\alpha\partial_x\theta_1+\sin\alpha\partial_x\theta_2)
{\big ]},
\end{align}
where 

\begin{align}
\label{DS5}
v_1=\frac{v_\rho+v_c}{2}+\frac{v_\rho-v_c}{2}\sqrt{1+\frac{4v_{\rho c}^2}{(v_\rho-v_c)^2}};\nonumber\\
v_2=\frac{v_\rho+v_c}{2}-\frac{v_\rho-v_c}{2}\sqrt{1+\frac{4v_{\rho c}^2}{(v_\rho-v_c)^2}}.
\end{align}
One can check that $v_1\sim v_\rho$ and $v_2\sim v_c$.

We now find corrections to the tunneling exponent $g$ in the first order in $v_{\rho n}$. We first find the correction due to the contribution $C_1=\sqrt{2}v_{\rho n}\partial_x\phi_n\cos\alpha\partial_x\theta_1$ in the last line of the action (\ref{DS4}) and then the correction from the contribution $C_2=\sqrt{2}v_{\rho n}\partial_x\phi_n\sin\alpha\partial_x\theta_2$. The strategy is the same on both steps. In particular, during the first step we omit $C_2$ from the action and diagonalize the remaining action with the transformation 

\begin{align}
\label{DS6}
\phi_n=\Phi\cosh\gamma+\Theta\sinh\gamma;\nonumber\\
\theta_1=\Phi\sinh\gamma+\Theta\cosh\gamma,
\end{align}
where 

\begin{equation}
\label{DS7}
\tanh 2\gamma=-\frac{\sqrt{2}v_{\rho n}\cos\alpha}{v_1+v_n}.
\end{equation}
This yields the action that consists of three pieces: one depends only on $\theta_2$, the second depends only on $\Phi$ and the third depends only on $\Theta$. We use that action to find the scaling dimensions of the quasiparticle operators 

\begin{equation}
\label{DS8}
\hat T=\exp\left(i\frac{\cos\alpha}{2\sqrt{2}}\theta_1+i\frac{\sin\alpha}{2\sqrt{2}}\theta_2\pm i\frac{\phi_n}{2}\right).
\end{equation}
This yields $g=\frac{3}{8}\mp\frac{v_{\rho n}\cos^2\alpha}{2(v_1+v_n)}$. Combining the corrections due to both contributions to the last line of Eq. (\ref{DS4}) one finds

\begin{equation}
\label{DS9}
g=\frac{3}{8}\mp\frac{v_{\rho n}}{2}\left(\frac{\cos^2\alpha}{v_1+v_n}
+\frac{\sin^2\alpha}{v_2+v_n}
\right).
\end{equation}
By inspecting different allowed relations between the parameters of the action (\ref{DS1}), one finds that $g\approx 3/8$ in all cases.

%\pagebreak
%\newpage


\begin{thebibliography}{99}
\bibitem{footnote} Disorder may lead to the formation of multiple islands, surrounded by edge channels of a width $a_0\ge l_B$ and separated by a distance $s$, $a_0\ll s\ll a$. In such scenario, the expression for $v_n$ contains the factor $sl_B/(aa_0)$ in place of $l_B/a$.
\end{thebibliography}
\end{document}